\documentclass[fleqn,usenatbib]{mnras}

\usepackage{newtxtext,newtxmath}
\usepackage{makecell}
\usepackage{longtable}
\usepackage[T1]{fontenc}

\DeclareRobustCommand{\VAN}[3]{#2}
\let\VANthebibliography\thebibliography
\def\thebibliography{\DeclareRobustCommand{\VAN}[3]{##3}\VANthebibliography}

\usepackage{graphicx}	% Including figure files
\usepackage{tikz,xcolor,hyperref}

\definecolor{lime}{HTML}{A6CE39}
\DeclareRobustCommand{\orcidicon}{%
	\begin{tikzpicture}
	\draw[lime, fill=lime] (0,0) 
	circle [radius=0.16] 
	node[white] {{\fontfamily{qag}\selectfont \tiny ID}};
	\draw[white, fill=white] (-0.0625,0.095) 
	circle [radius=0.007];
	\end{tikzpicture}
	\hspace{-2mm}
}

\foreach \x in {A, ..., Z}{\expandafter\xdef\csname orcid\x\endcsname{\noexpand\href{https://orcid.org/\csname orcidauthor\x\endcsname}
			{\noexpand\orcidicon}}
}

\title[Mass of Open clusters]{Revisiting the mass of open clusters with \emph{Gaia} data}

\author[A. Almeida et al.]{
Anderson Almeida,$^{1}$\thanks{E-mail:andersonalmeida\_sa@outlook.com}
Hektor Monteiro$^{1\orcidB{}}$,
Wilton S. Dias$^{1}$
\\
$^{1}$Instituto de F\'isica e Qu\'imica, Universidade Federal de Itajub\'a, Av. BPS 1303 Pinheirinho, 37500-903 Itajub\'a, MG, Brazil\\
}

\date{Accepted XXX. Received YYY; in original form ZZZ}

\pubyear{2023}

\begin{document}
\label{firstpage}
\pagerange{\pageref{firstpage}--\pageref{lastpage}}
\maketitle

\begin{abstract}

The publication of the \emph{Gaia} catalogue and improvements in methods to determine memberships and fundamental parameters of open clusters has led to major advances in recent years. However, important parameters such as the masses of these objects, although being studied mostly in some isolated cases, have not been addressed in large homogeneous samples based on \emph{Gaia} data, taking into account details such as binary fractions. Consequently, relevant aspects such as the existence of mass segregation were not adequately studied. Within this context, in this work, we introduce a new method to determine individual stellar masses, including an estimation for the ones in binary systems. This method allows us to study the mass of open clusters, as well as the mass functions of the binary star populations. We validate the method and its efficiency and characterize uncertainties using a grid of synthetic clusters with predetermined parameters. We highlight the application of the method to the Pleiades cluster, showing that the results obtained agree with the current consensus in the literature as well as recent \emph{Gaia} data. We then applied the procedure to a sample of 773 open clusters with fundamental parameters determined using \emph{Gaia Early Data Release 3 (eDR3)} data, obtaining their masses. Subsequently, we investigated the relation between the masses and other fundamental parameters of the clusters. Among the results, we found no significant evidence that clusters in our sample lose and segregate mass with age. 

\end{abstract}

\begin{keywords}
Galaxy: open clusters and associations: general
\end{keywords}

%%%%%%%%%%%%%%%%%%%%%%%%%%%%%%%%%%%%%%%%%%%%%%%%%%

%%%%%%%%%%%%%%%%% BODY OF PAPER %%%%%%%%%%%%%%%%%%

\section{Introduction}

Open clusters are an interesting class of objects to investigate the Galaxy structure and evolution, since their distances and ages can be determined with good precision. Young open clusters can be used to investigate and understand the physics of star formation processes (see \citet{Krumholz2019} and \citet{Krause2020} for a review) and the relationship of mass to other parameters of those objects, such as distance and age, allows us to understand how they evolve dynamically over time. These properties also help to constrain the initial conditions used in N-body simulation models which investigate cluster evolution \citep{Pijloo2015, Rossi2016}.

Clusters are affected and disrupted by internal processes (stellar winds, internal interactions, mass loss due to stellar evolution) and external processes (supernova, collision with molecular clouds, forces interacting with the Galactic potential) depending on their lifetime. These processes could change the mass of the clusters and reduce their size \citep{Spitzer1958,Joshi2016} although the details are still being debated. Clusters may also dissolve with dissolution times that can depend on several factors such as initial cluster mass and internal structure, as well as on the external effects mentioned previously (see for example \citet{Goodwin2006} and \cite{Portegies2010} regarding massive young clusters).

The publication of data from the \emph{Gaia} satellite has led to an increased number of open clusters for which fundamental parameters have been determined \cite[see][for a review]{Cantat2022}. While the estimation of their mass remains challenging, various techniques have been employed using kinematic or photometric data. An important method is based on the density profile of the cluster, which is described by the King profile \citep{perfilking1962}. The King profile method involves fitting a King profile function to the measured density profile of the cluster, assuming that the cluster is in dynamical equilibrium. The determined core radius and tidal radius of the cluster are then used to calculate the mass. Direct use of the Virial theorem has also been used to get the cluster mass, as discussed in \citet{McNamara1983} and \citet{ Raboud1996}. Integrated stellar luminosity function is another widely used method involving the construction of a histogram of the number of stars as a function of their luminosities and comparing it to theoretical models. Using photometric data, one of the first works to apply this strategy was \citet{vandenBergh1984} based on a sample of 142 clusters. In, \citet{Piskunov2008LF} for example, the authors use the method to study the initial mass function of open clusters. However, for all strategies mentioned, lack of precise knowledge of the cluster member stars as well as their binary fractions were important sources of uncertainties.

In the pre-\emph{Gaia} years, some results on the structure and dynamics of clusters are noteworthy due to the quality of the data and/or the method used. Below, we highlight a non-exhaustive selection of studies relevant to the present work.

Due to its proximity, Pleiades is one of the most studied open clusters, however, only in 1998 with the work of \citet{Raboud1997} it was possible to reliably estimate its mass based on a list of 270 members. The authors found mass segregation in binary and individual stars with masses up to 1.0 ${M}_\odot$. They estimated the mass of the Pleiades through different methods: tidal radius $(4000 \pm 4000)M_{\odot}$, Virial theorem $(720 \pm 220)M_{\odot}$, mass function integration $(950 \pm 200)M_{\odot}$ and by the sum of the stellar masses of the sample $(412)M_{\odot}$.

In \citet{Raboud1998} the authors studied the radial structure of Praesepe and of the very young open cluster NGC 6231. They considered the structure and dynamics of the clusters using mass estimates of individual member stars. In both clusters, the masses of the stars were estimated only for the main sequence by means of a visual isochrone fit. The authors found mass segregation among the cluster members with a less pronounced degree of segregation in Praesepe compared to Pleiades: a surprising fact according to the authors, as ancient clusters, such as Praesepe, were expected to have a greater degree of segregation. The mass of Praesepe was estimated to be (590) ${M}_\odot$ by integrating the mass function.

Only few works present results of mass for dozens of clusters and those were mainly based on the data from 2MASS catalogue \citep{Skrutskie2006} and stellar membership from the ground-based astrometric PPMX catalogue \citep{PPMX2008} published in MWSC \citep{catalogoKharchenko2013}. \citet{extrapolacaobr} determined the mass for 11 nearby open clusters, taking into account the unobserved main sequence stars by extrapolating the mass functions to the low mass limit of $0.08$$M_{\odot}$.  They also included estimates of the mass of unobserved evolved stars. They analyzed mass segregation, the dynamic evolution of clusters, and its implications for the slopes of mass functions, reporting that cluster size correlates to age, Galactocentric distance, and mass. \citet{PiskunovClusters2007} and \citet{PiskunovTidal2008} derived the tidal radii and masses of 236 clusters, selected from the ASCC-2.5 catalogue, \citep{catalogoASCC2.5} by fitting King profiles \citep{perfilking1962}. They found that the tidal radius distribution had peaked at about 1.5 pc and 7-10 pc and that most clusters have tidal masses in the range $log(M_{c}/M_\odot)$ = 1.6 and 2.8. \citet{Bukowiecki2011b} estimated the mass of a sample of 599 clusters, converting the luminosity function into the mass function using the fitted isochrone. 

\citet{Joshi2016} studied 1241 open clusters, with distances of up to 1.8 kpc from the Sun. Based on data from the MWSC \citep{catalogo2013_kharchenko} catalogue and masses from \citet{PiskunovTidal2008}, the work presents statistical analyses of several fundamental parameters such as spatial position, age, size, mass, and extinction. They found that massive clusters with larger radii, within the solar orbit, tend to have better chances of surviving external interactions and that younger clusters have larger masses and tend to lose mass at rates of about $(150)M_\odot$/Myr.

With the availability of \emph{Gaia} data, some works have emerged that, among other aspects, address the cluster mass estimates. For example, \citet{Yontan2019} studied the clusters ASCC 115, Collinder 421, NGC 6793, NGC 7031, NGC 7039, NGC 7086, Roslund 1, and Stock 21 where combining CCD UBV photometric and \emph{Gaia} astrometric data showing that the slopes of the mass functions of the clusters are in good agreement with the value of \citet{Salpeter1955...121..161S}. In the work of \citet{Rangwal2019}, three clusters were analysed: NGC 6067, NGC 2506, and IC 4651. The photometric data for the clusters were converted into mass functions using theoretical models. Similar results for IC 4651 were also observed in \citet{Pandey1992} and \citet{Durgapal2001} where the existence of mass segregation was suggested.

Using fractal statistics, \citet{Hetem2019} investigated a sample of 50 young open clusters  in different stages of evolution at different regions of the Galaxy. The individual stellar masses were obtained by comparing the observed color-magnitude diagram with theoretical evolutionary tracks. The authors estimated the mass of the clusters by the sum of the individual mass of the member stars, which were determined from \emph{Gaia DR2} astrometric data \citep{GAIA-DR22018}. Their results indicate that 46\% of the clusters did not show mass segregation and the majority of the objects were inconclusive due to large errors, with only 4 objects showing clear mass segregation.

Combining data from \emph{Gaia DR2}, 2MASS, WISE, APASS and Pan-STARRS1, \citet{Bisht2020} studied the open clusters Czernik 14, Haffner 14, Haffner 17, and King 10 and estimated their mass by integrating their mass function and concluded that all clusters are dynamically relaxed and show mass segregation.

More recently, two works have studied masses of open clusters. First, \cite{masstotal46ocs2021}, using the \emph{Gaia eDR3} catalogue, studied 46 clusters with relatively elongated morphology to determine their tidal radii and study their disintegration. In addition, they determined the initial masses of the clusters and estimated their total observed masses using mass function integration. The authors concluded that 41 objects had some degree of mass segregation. Second, \cite{masstotalEbrahimi2022} investigated 15 nearby clusters using \emph{Gaia DR3} data. The authors use synthetic clusters, comparing their stellar populations with the observed ones to determine their current mass functions. From these mass functions, they estimate the masses and the binary fractions. They find that the obtained mass functions were consistent with a single power-law function. Also, a significant correlation between the mass function slope and the ratio of age to half-mass relaxation time was found. Their results also indicate that the less evolved objects have a mass function consistent with that of the solar neighbourhood, indicating a possible connection between the dissolution of open clusters and the formation of the Galactic disc.

In this study, we re-examine the potential correlations of open cluster masses with their ages, radius, and mass segregation, using a substantial sample of objects. We obtain updated estimates of mass using the \emph{Gaia eDR3} catalogue \citep{GaiaEdr3collaboration} in order to investigate these relationships. In Sec. 2 we present a new procedure to obtain mass estimates of individual cluster member stars, taking into account binaries and, from those, the total cluster masses using the present-day mass functions. The section also presents an extensive validation of the results based on a set of synthetic clusters. In Sec. 3 the procedure is applied to the open cluster Pleiades, showing that the results obtained agree with others in the literature, in particular recent \emph{Gaia DR3} mass data. In Sec. 4 we apply the method to a sample of 773 real open clusters and discuss the results, also comparing to previous literature values when available. In Sec. 5 we give our conclusions.

%\footnote{Table \ref{tab:totalMasses} and the cluster members are only available in electronic form at the CDS via anonymous ftp to \url{http://cdsarc.cds.unistra.fr} or via \url{https://cdsarc.cds.unistra.fr/cgi-bin/qcat?J/A+A/}

\section{Methodology}

In this study, we propose a new method to compute the masses of open clusters, here considered as the mass of member stars, which were determined through membership methods. The method is based on the determination of individual cluster member star masses using precise age, distance, metallicity, and extinction parameters obtained from \emph{Gaia} data. To estimate the individual member star masses, synthetic clusters are generated with full control of fundamental parameters. Finally, a Monte Carlo method is used to obtain the final individual masses, as well as their uncertainties, by comparing the observed and generated synthetic clusters. From the individual masses of the stars and an estimation of the unseen mass in low mass stars as well as remnants of evolved stars, the masses of the clusters are obtained. We validate the method by using a set of simulated observations based on synthetic clusters generated from predefined isochrones of given age, distance, extinction and metallicitiy as well as binary fraction.

The procedure is summarized in the steps below:

\begin{itemize}
      \item the fundamental parameters and photometric data are obtained for a sample of open clusters from \citet{catalogoDias2021}, with updated values based on \emph{Gaia eDR3} data;
      
      \item the fundamental parameters are used in the generation of synthetic clusters which will be compared to the observed one in a Monte Carlo procedure;
      
      \item the Monte Carlo procedure generates a sample of individual mass estimates for each observed cluster star;

      \item based on comparison to the synthetic clusters, stars are labeled as binary or not and a companion mass is estimated if appropriate;

      \item the final individual star masses (and companions when present) are obtained by averaging the sample of mass estimates generated by the Monte Carlo procedure;

      \item the binary fraction and average mass fraction of binary systems are obtained from the observed stars;

      \item the final uncertainties are obtained by the standard deviation of the sample of mass estimates;
      
      \item after calculating individual star masses, we obtain the present-day mass function by fitting a function to the cluster stellar mass distribution obtained via a histogram of the mass estimates;

      \item the fitted present-day mass function is subsequently used to extrapolate the population of unseen low-mass stars, as well as the unseen mass in remnants of evolved stars;

      \item the mass obtained for the observed stars (single and binary), the mass of unseen low mass stars, the mass of remnants of evolved stars and the contribution of unseen binaries are added to obtain the final mass of the cluster.
\end{itemize}

The procedure is described in more detail in the following sections.

\subsection{Determination of stellar masses}

\begin{figure}
\begin{center}
\includegraphics[width=\columnwidth]{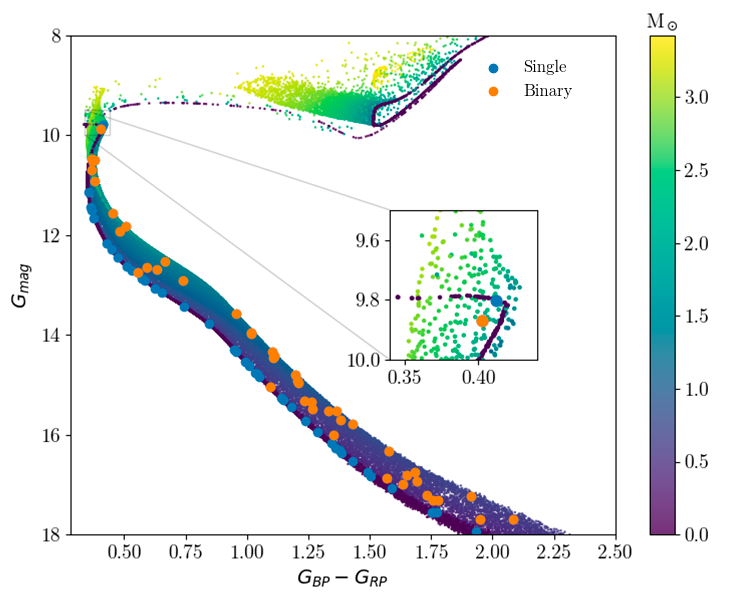}
\caption{CMD of a simulated open cluster of log(age) = 8.5, distance = 1.0 kpc and $A_v$ = 1.0 mag showing 50 single and binary member stars overlaid on a synthetic cluster generated with the same referenced parameters consisting of 10000 synthetic stars. The stars in the synthetic cluster are coloured based on the mass of the companion where the main sequence is set to $M_{comp} = (0.0)M_{\odot}$. The zoom in the figure shows where the main sequence region overlaps with the binary star region.}
\label{fig:sobreposicoes}
\end{center}
\end{figure}

\begin{figure}
\begin{center}
\includegraphics[width=\columnwidth]{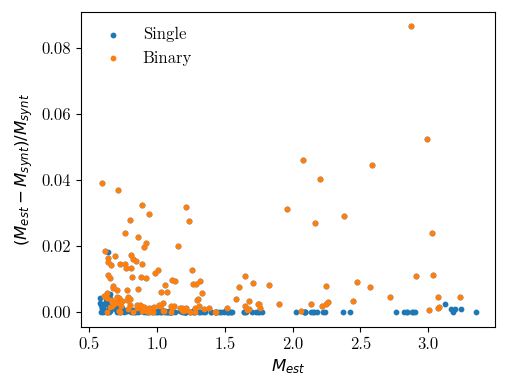}
\caption{Test results using our mass determination method performed on a synthetic cluster of log(age)=8.5, distance = 1.0 kpc and $A_v$ = 1.0 mag. The relative error based on the mass estimate ($M_{est}$) and the expected synthetic mass ($M_{synt}$) is presented for single stars (blue) and binary stars (orange).}
\label{fig:mass_error}
\end{center}
\end{figure}

In our procedure, the estimation of the mass of each member star observed in a given open cluster is the fundamental necessary step that needs to be carried out, before the cluster mass can be obtained. As mentioned previously, with the open cluster member stars defined, the Monte-Carlo method compares generated synthetic clusters to the observed ones to determine the masses of the individual member stars. To generate the synthetic cluster we need a given set of theoretical isochrones and in this work we use the Padova PARSEC version 1.2S set of isochrones described in detail in  \citet{Bressan2012}\footnote{available at \url{http://stev.oapd.inaf.it/cgi-bin/cmd}}. The synthetic clusters are generated with a predefined number of stars with masses drawn from a probability distribution parameterized by the well-known Chabrier initial mass function \citep{Chabrier2003}. In this work, we adopted a number of generated stars of 10000. A fraction of binary stars of 50\% is also adopted, with the binary system companion star masses drawn from a uniform distribution with values varying from 0 to the primary star mass. Photometric errors, as per the \emph{Gaia} data definition \citep{Gaiacolab2016}, are also included in the synthetic cluster.

For each star in the observed cluster, we compared its magnitudes to each of the 10000 stars generated in the synthetic cluster, finding the one with the smallest difference in magnitude. This step effectively finds the synthetic star with the minimum Euclidean distance from the observed one in magnitude space, according to the following equation:

\begin{equation}
d_i  = \min_{s\in S} ~ \sqrt {\sum _{j=1}^{m}  \left( O_{ij}-S_{sj}\right)^2 }.
\end{equation}    

where $d_i$ is the minimum distance of the star of the observed cluster $O$, to the $s$ stars in the synthetic cluster $S$, measured in the $jth$ magnitude band of the $m$ distinct bands available (\emph{Gaia eDR3}: G, GBP and GRP in this work).

With the closest synthetic star determined, we check if it corresponds to a binary. If it is, we mark the observed star as a binary and assign to it the corresponding primary and secondary masses, since for the synthetic cluster those are known. Else, we assign the mass of the nearest synthetic cluster star to the observed star. The process is repeated a large number of times, each time with a distinct synthetic cluster generated by varying individual stars. The final mass of the observed stars and their companions are adopted as the median of the respective sample of masses determined. The errors are obtained by the standard deviation of the estimates in each case. We note here that, since the method determines masses based on the minimization of Euclidean distances in magnitude space, the distribution of binary system companion masses is adopted for the synthetic cluster solely to efficiently sample the CMD region where binaries are likely to be located. The actual distribution of mass fractions in binary systems for a given observed cluster is determined a posteriori from the estimated stellar masses. The procedure is illustrated in Fig. \ref{fig:sobreposicoes} where we have generated a simulated open cluster with 50 member stars overlaid on a synthetic cluster generated with the same referenced parameters with 10000 synthetic stars.

To validate the stellar mass estimation method, a grid of synthetic clusters was generated with log(age) varying from 6.6 to 9.5; reddening from 0.5 to 3.0 magnitudes; distances from 1 to 5 kpc and with 300 observed stars, 50\% of which are binary systems. We then compared the results of our mass estimation procedure to the input masses of the generated grid. In Fig. \ref{fig:mass_error} we show an example of the validation for a simulated cluster of log(age) = 8.5, distance = 1.0 kpc, and $A_v$ = 1.0 mag. We can see that results are in good agreement with the input synthetic masses, with binary systems showing larger differences as expected.

\begin{figure}
\begin{center}
\includegraphics[width=\columnwidth]{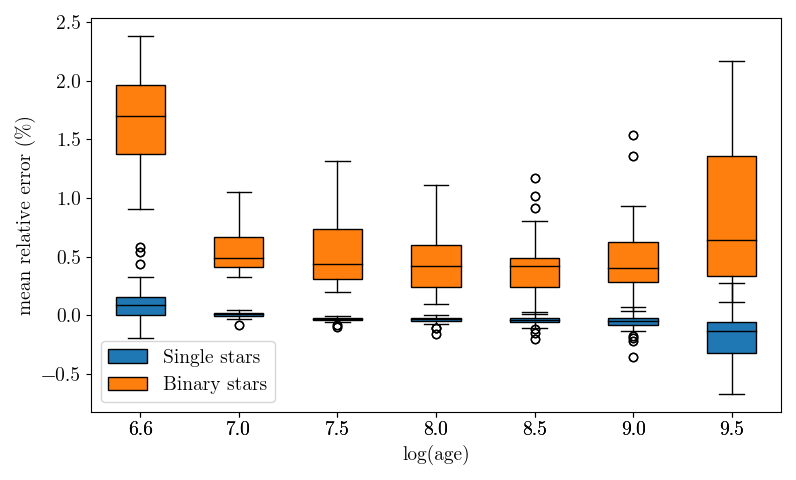} \\
\includegraphics[width=\columnwidth]{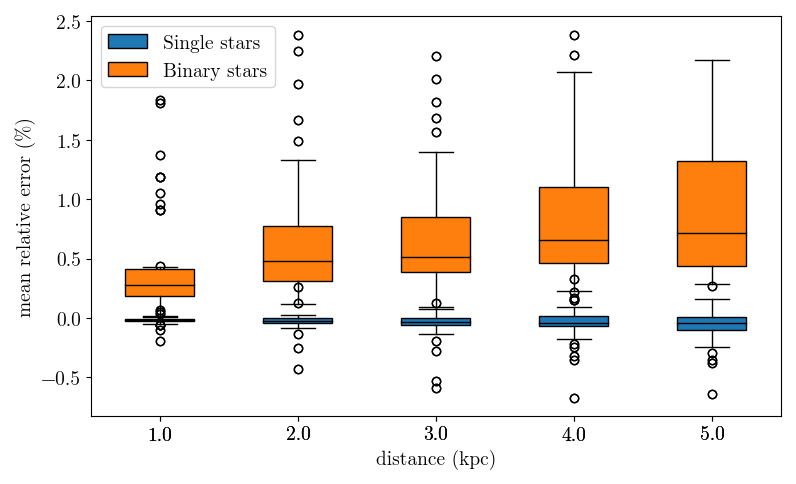}
\caption{Mean relative error committed in the estimate of individual stellar masses as a function of the cluster age and distance.}
\label{fig:indMass_error}
\end{center}
\end{figure}

The individual mass estimates also vary with respect to cluster ages and distances, as shown in Fig. \ref{fig:indMass_error}. As expected, the relative errors are worst for binaries and increase with age and distance, also with increasing spread. The behaviour is expected as the photometric errors affect proportionally more the clusters at higher distances. Young clusters also show larger uncertainties due to their more undefined turn-off region in colour-magnitude space.

\begin{figure}
\begin{center}
\includegraphics[width=\columnwidth]{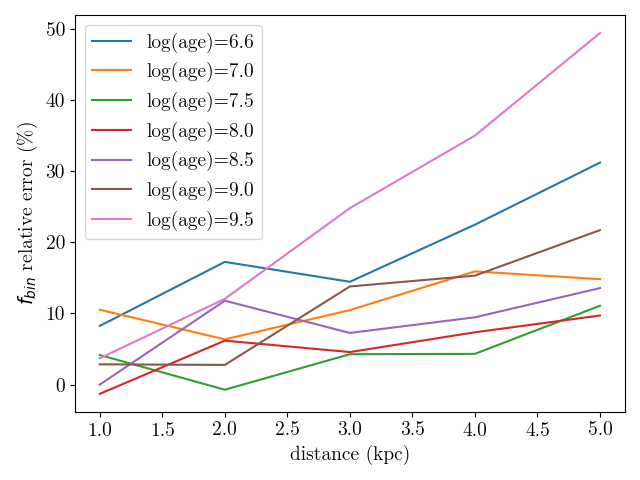} \\
\includegraphics[width=\columnwidth]{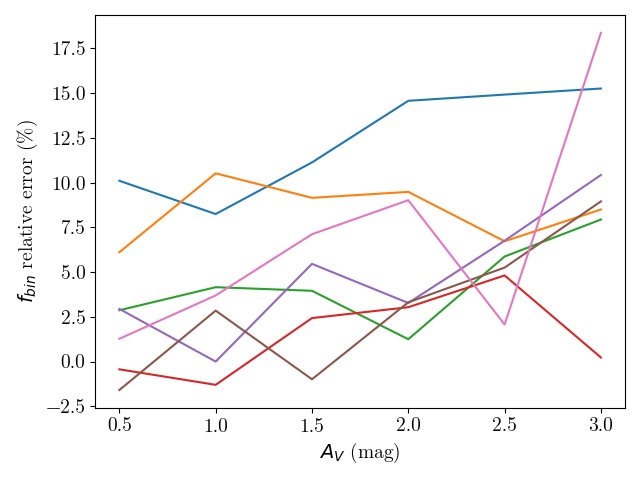}
\caption{Relative error of the binary fraction ($f_{bin}$) found by the mass estimation procedure considering the known fraction (adopted as 50\% of binaries) in simulated clusters. In the upper panel where the dependence to distance is investigated, we set $A_V=1.0$ and in the lower panel where the dependence to $A_V$ is investigated we set distance to 1kpc.}
\label{fig:teste_binarias}
\end{center}
\end{figure}

We tested the efficiency of the binary detection feature in relation to other parameters of the simulated clusters and found some relevant systematic effects as a function of distance and of extinction $A_V$. In Fig. \ref{fig:teste_binarias} we show the result of comparing the fraction of binaries found by the mass estimation method to the known binary fraction, adopted as 50\% for the whole grid of simulated clusters. We can see a clear overestimation of the binary fraction with the increase of both distance and extinction, as well as age. For clusters with log(age) < 9.0 and distance smaller than about 2.5 Kpc, the overestimation is about 20\%. This effect is a result mainly of the increased uncertainty in the fainter magnitudes and the smaller sampling of the main sequence region for clusters at higher ages and distances.

Despite the good accuracy and precision shown in our method, we should note some limitations. In particular, detecting binary stars is a problem in regions of the CMD where the main sequence and binary star regions overlap. In Fig. \ref{fig:sobreposicoes} we present a plot of a CMD of a simulated open cluster of log(age) = 8.5, distance = 1.0 kpc and $A_v$ = 1.0 mag showing 50 single and binary member stars overlaid on a synthetic cluster generated with the same referenced parameters consisting of 10000 synthetic stars. Parallel to the main sequence, the region of binary stars can be seen as well as the effect of the varying companion masses, such as the horizontal sequence at about G $\sim$ 9. The zoom in the figure shows the main sequence region where an overlap with the binary star region occurs at about G $\sim$ 10. It is very difficult to define a constraint to determine a correct mass value in these regions of the CMD. Fortunately, this problem occurs in small regions only. Also, in most cases, the mass estimation error is usually small and does not show systematic trends, as can be seen in Fig. \ref{fig:mass_error}.

%%%%%%%%%%%%%%%%%%%%%%%%%%%%%%%%%%%%%%%%%%%%%%%%%%%%%%%%%%%%%%%%%%%%%%%%%%%%%%%%

\subsection{Determination of cluster masses}
\label{sec:methods}

With the individual masses of the member stars of a given observed cluster, we can obtain its mass by considering the observed mass and an estimation of the unobserved mass. To estimate the unobserved mass in low mass stars, we obtain the present-day mass function from the observed star masses and extrapolate it to the lower mass limit of 0.09$M_{\odot}$, adopted based on the limit of the isochrones used. We also estimate the unobserved mass of evolved stars by extrapolating in the high mass interval, if necessary, and obtaining the number of stars that have evolved to white dwarfs. In our work, we assume that the mass function can be represented by a two-part segmented linear function, which is then fit to the mass distribution of observed stars and extrapolated to the lower mass limit. The two-part segmented linear function is defined as:

\begin{equation}
    f(x) = \begin{cases}
    \alpha_Bx + b_1, & \text{if } x \leq M_c \\
    \alpha_Ax + b_2, & \text{if } x > M_c
\end{cases}
\end{equation}

where $\alpha_A$ and $\alpha_B$ represent the slopes of the function for the high mass slope and low mass slope, and $b_1$ and $b_2$ are the corresponding intercepts, respectively. $M_c$ is the mass value at which the transition between the two linear segments occurs. In our fitting procedure $\alpha_A$, $\alpha_B$ and $M_c$ are free parameters and $b_1$ and $b_2$ are taken such that $f(M_c^{-}) = f(M_c^{+})$.

The segmented linear function is fitted to a histogram for the mass distribution, where the number of bins is obtained using the rule of \citet{sturges}. The fits were performed using the routine \textit{curve\_fit}\footnote{\url{https://docs.scipy.org/doc/scipy/reference/generated/scipy.optimize.curve_fit.html}} with the default Trust Region Reflective algorithm. The fits were done for the entire original sample\footnote{A sample of the sequence including the functions used is available at \url{https://github.com/ander-son-almeida/DashboardOCmass/blob/main/examples/MF_example.ipynb}} of 1743 open clusters from \citet{catalogoDias2021}. We then evaluated each result to select only those that satisfied the requirement that the histogram for the mass distribution had at least 2 bins on both sides of the fitted transition mass point $M_{c}$. The final cluster sample consists of 773 objects that satisfied this criterion.

The unobserved mass of evolved stars is obtained by estimating how many of the cluster stars have evolved to white dwarfs, which are typically not detected due to their low luminosity. To achieve this, we use the initial to final mass relation for progenitor stars from 0.85 to 7.50$M_{\odot}$ provided by  \citet{Cummings2018}. From the fitted mass function, we calculate the number of stars in that mass range and then use the relation to obtain the present-day mass in white dwarfs.

The unobserved mass in the low mass range is obtained by integrating the mass function from the lowest observed mass to the low mass limit of 0.09$M_{\odot}$. The unobserved low mass range is defined by low luminosity stars not detected by \emph{Gaia}, or stars with magnitudes greater than 19, which were removed from our sample due to Gaia's incompleteness limit. This procedure is similar to the one adopted by \citet{extrapolacaobr}.

It is important to note that in this step, we do not include the masses of the binary companion to construct the present-day mass function. Since we do not know the distribution of binaries in the unobserved portion, we estimate their contribution to the mass by first obtaining a mean mass fraction for binary systems, given by $q = \overline{M_{secondary}/M_{primary}}$, from the observed stars. We then obtain the binary system mass from the estimate of primary masses in the unobserved mass interval. For these systems, we assume that the primaries are distributed according to the mass function determined by the observed stars. The total unseen mass of primaries is estimated by integrating the extrapolated observed mass function down to the lower mass limit. The contribution of secondaries is found by multiplying this value by the estimated mean mass fraction to obtain the masses of the companions, thus giving the final mass in binaries. The unobserved mass will be a sum of the mass of single stars and binary systems down to the lower mass limit.  The binary fraction used is also determined from the observed stars. The final mass of the cluster is then estimated by:

\begin{equation}
    M_T = M_o + M_{un} (1+F_{bin} \times q) + M_{WD}
\end{equation}

where $M_T$ is the mass of the cluster, $M_o$ is the mass estimated from observed stars, $M_{un}$ is the mass obtained from the integrated mass function for unobserved stars, $F_{bin}$ and $q$ are the estimated fraction of binaries and mean mass ratio in binary systems respectively, estimated from the observed stars and $M_{WD}$ is the estimated mass of white dwarfs.

\begin{figure}
\begin{center}
\includegraphics[width=\columnwidth]{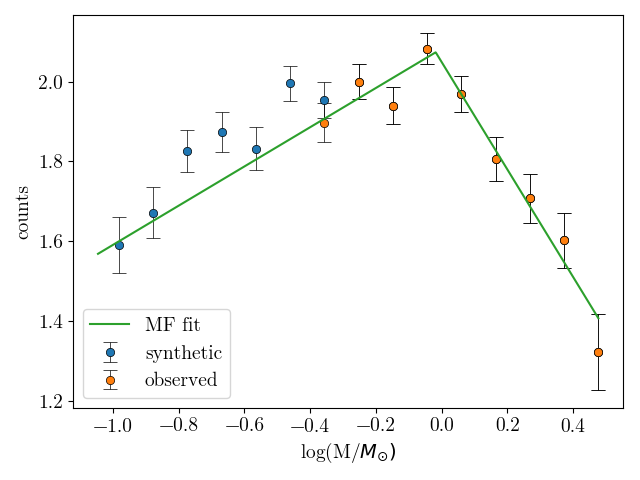}
\caption{Synthetic open cluster of log(age) = 8.5, distance= 0.5 kpc, $A_v$ = 1.0 mag, [Fe/H] = 0.0, the mass of 1043$M_{\odot}$ and the respective segmented line fit to the simulated observation (orange dots). The estimated unseen mass is the region defined by the blue dots, where the fitted function is used to obtain the integrated mass.}
\label{exemplo_extrapolacao}
 \end{center}
 \end{figure}

\begin{figure*}
\centering
{\includegraphics[width=\columnwidth]{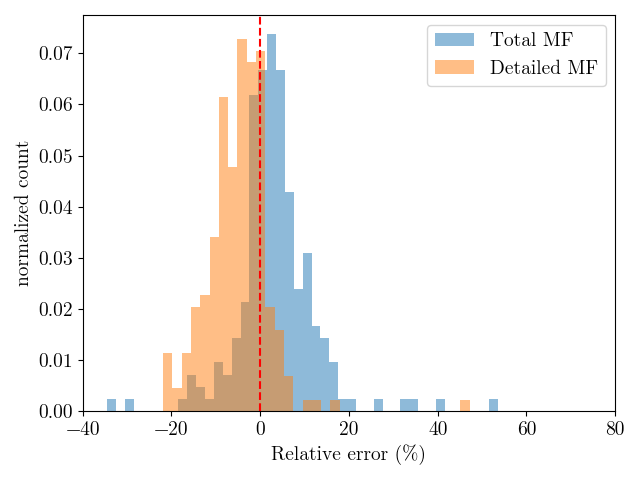}}
{\includegraphics[width=\columnwidth]{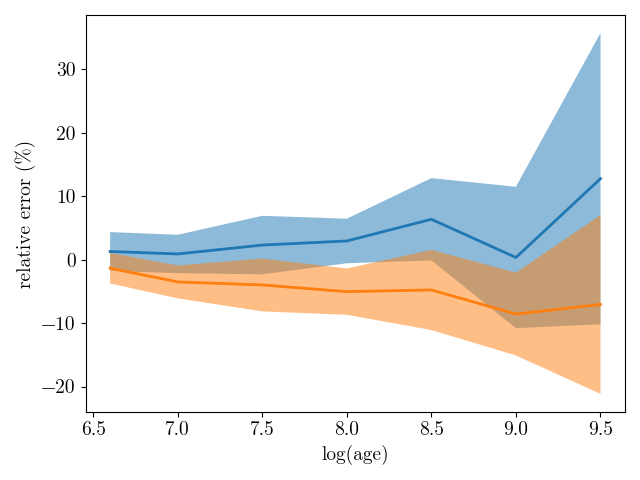}} \\
{\includegraphics[width=\columnwidth]{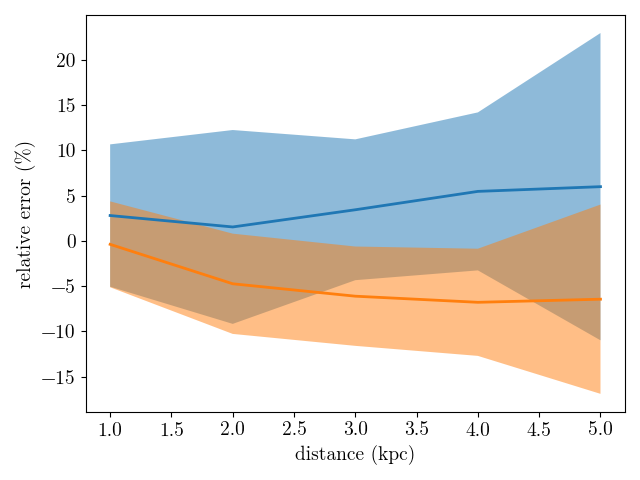}}
{\includegraphics[width=\columnwidth]{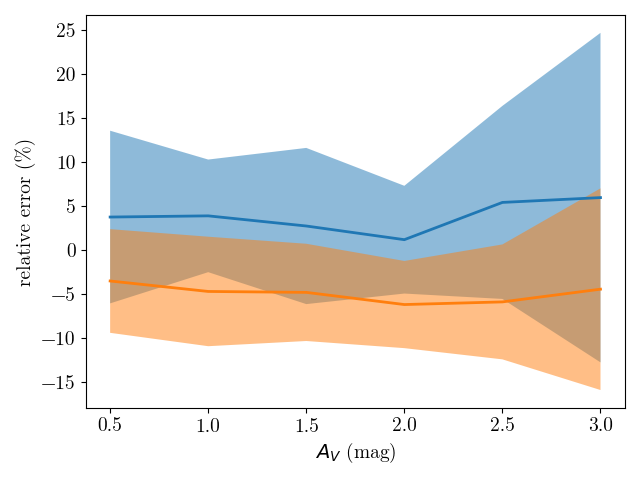}} \\
\caption{Distribution of the relative error on the cluster's mass obtained for each method used (upper left), average relative error and the respective 1$\sigma$ uncertainty as a function of log(age) (upper right), distance (lower left) and $A_V$ (lower right).}
\label{fig:erro_massa_total}
\end{figure*}

To exemplify the extrapolation of the unobserved region of low-mass stars, in Fig. \ref{exemplo_extrapolacao} we show a mass function obtained for a simulated open cluster with 600 stars with log(age)=8.5, at a distance of 500pc, $A_V$=1.0, [Fe/H]=0.0,  the mass of 1043$M_{\odot}$. The segmented line fit (green line) to the simulated observation (orange dots) is also shown. We see from the figure that the segmented line function fits well to the higher mass end of the mass function while less to the lower mass section, especially the unobserved low mass region (blue dots) which is generated from a Chabrier \citep{Chabrier2003} initial mass function. The mass obtained with the fit to the integrated mass function in this example was 1071$M_{\odot}$, in agreement with the mass of 1043$M_{\odot}$ of the simulated cluster.

The \emph{Gaia eDR3} data quality has improved significantly the definition of open cluster members and features in the CMDs, enabling the determination of mass functions for individual, primary, and secondary stars in some cases. Consequently, we were able to investigate two methods to calculate the masses for a subsample of the open clusters:

\begin{itemize}

\item {\bf mass function}: the determination of the mass of a cluster as described previously, where we obtain a mass function with all the observed stars, excluding binary companions, and from that, estimate the unseen mass including the contribution of binaries;

\item {\bf detailed mass function}: the determination of the mass of the cluster by summing the observed star masses to the contribution of unobserved stars as determined from mass functions determined for single, primary and secondary stars separately.

\end{itemize}

To validate both procedures and to investigate their limitations, we again make use of the grid of synthetic clusters with different parameters. In Fig. \ref{fig:erro_massa_total} we present the distribution of the relative error of the mass for each method used and the error as a function of the age, distance and extinction $A_V$. The results indicate that the mass error tends to increase with cluster age, distance and extinction being usually under 20\%. This behavior is expected since more distant and old open clusters present a low sampling of the main-sequence stars and suffer from larger photometric uncertainties. There is also a small systematic overestimation of the mass by the integrated mass function method and a small underestimation in the detailed mass function method. Since it is computationally difficult to obtain individual mass error estimates for each cluster studied, we will adopt the conservative 20\% value overall as our uncertainty in this work.

%%%%%%%%%%%%%%%%%%%%%%%%%%%%%%%%%%%%%%%%%%%%%%%%%%%%%%%%%%%%%%%%%%%%%%%%%%%%%%%%

\section{Pleiades: a benchmark cluster}

With the method tested and validated in a large grid of synthetic clusters, we proceeded to apply it to the Pleiades cluster, which can be considered a benchmark case since there are several studies on its stellar population and precise astrometric and photometric data. These data, especially from the \emph{Gaia} catalogue, have allowed a precise determination of member stars and features in the cluster CMD. We took the sample of 1236 astrometric members, determined individual masses and applied the two methods described in the previous section to determine the cluster mass. 

\begin{figure}
\begin{center}
\includegraphics[width=\columnwidth]{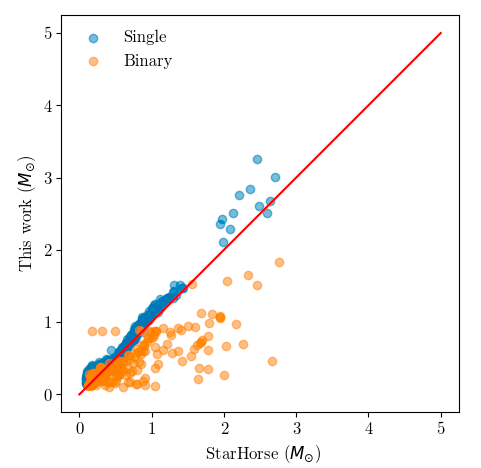} \\
\includegraphics[width=\columnwidth]{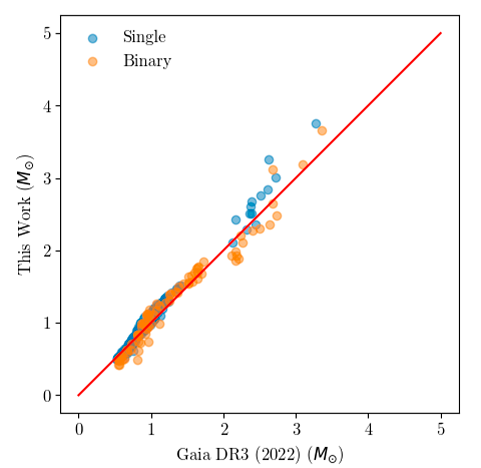}
\caption{Comparison of individual stellar mass results from our work with \citet{StarHorse} (upper panel) and \citet{GaiaDR3_mass} (lower panel). For the stars in binary systems the mass shown is that of the primary.}
\label{fig:StarHorse-gaia}
\end{center}
\end{figure}

We compared our individual mass estimates with the determination of \citet{StarHorse}, which used \emph{Gaia eDR3}, Pan-STARRS1, SkyMapper, 2MASS, and AllWISE photometric data. Although estimating the mass was not their main goal, they provide the values which we compare to our estimates in Fig. \ref{fig:StarHorse-gaia}, where it is possible to see a general good agreement but also a systematic deviation in the binary stars. It is important to point out that their method assumes stars are single, so the values estimated are relative to the star that dominates the photometry, usually the primary. This fact may explain the systematic underestimation of the primary mass value seen in their method.

Masses from \citet{GaiaDR3_mass} in data release DR3 were also available, and a comparison of the stars in common to our Pleiades member sample is shown in Fig. \ref{fig:StarHorse-gaia}. As in the previous case, we can see that our results are in good agreement, especially  single stars with masses lower than $\sim$ $2 M_{\odot}$. There are again some disagreements regarding the masses of the primary stars in binary systems, as \emph{Gaia} does not estimate the mass of the secondary stars in possible binary systems. However, here we see better agreement with our results with no clear systematic deviations.

As an example of individual mass estimate quality, in the observed stars, we found a mass of 4.94$M_\odot$ for the $20 Tauri$ star. The mass of this star determined by our method agrees with other values found in the literature, such as 4.22$M_\odot$ from \citet{2estrelamassivapleiades}, 4.29$M_\odot$ from \citet{1estrelamassivapleiades} and 4.90$M_\odot$ from \citet{3estrelasmassivaPleiades}.

It is also possible to check the consistency of the binary fraction estimated by our method. From the sample of the Pleiades member stars we identified, 1006 single stars and 230 binary systems, giving a binary fraction of 19\%. The binary fraction estimates found published in the literature ranges from 17\% to 50\% with many works focusing on the lower mass population \citep{1982AJ.....87.1507S, binarias1995MNRAS.272..630S, binariaspleiadesmartin2003hubble, binariaspleiades2007MNRAS.380..712L, Pinfield2003_2mass, Maxted2005, bate2002}. One of the first works to analyse the frequency of binaries in the Pleiades was \cite{1975PASP...87..707B}, which found a 22\% binary fraction for main sequence stars. Recently, \cite{torresbinaroas2020ApJ...901...91T} found a fraction of multiple and binary systems of 37\% (12/32), considering a list of 32 stars of spectral type B and A with data from \emph{Gaia DR2}. A study of spectroscopic monitoring \cite{binariaspleiades2021ApJ...921..117T} found a fraction of 17\% (48/289) which, after corrections for incompleteness, took on a value of ($25 \pm 3$)\%. Our binary fraction estimate is within the reported range and should be considered as a lower limit since the contamination of false positive detection of members in the low magnitude interval (G>16) may be significant.

From the masses of the primary and secondary stars of a binary system, we were able to calculate the average mass ratio in binary systems. For the binary stars of Pleiades, we found an average mass ratio of $q = \overline{M_{secondary}/M_{primary}} = (0.58 \pm 0.23)$. This value is close to that found by \cite{razao_binarios2022}, who determined ($0.73 \pm 0.03$).

\begin{figure}
\begin{center}
\includegraphics[width=\columnwidth]{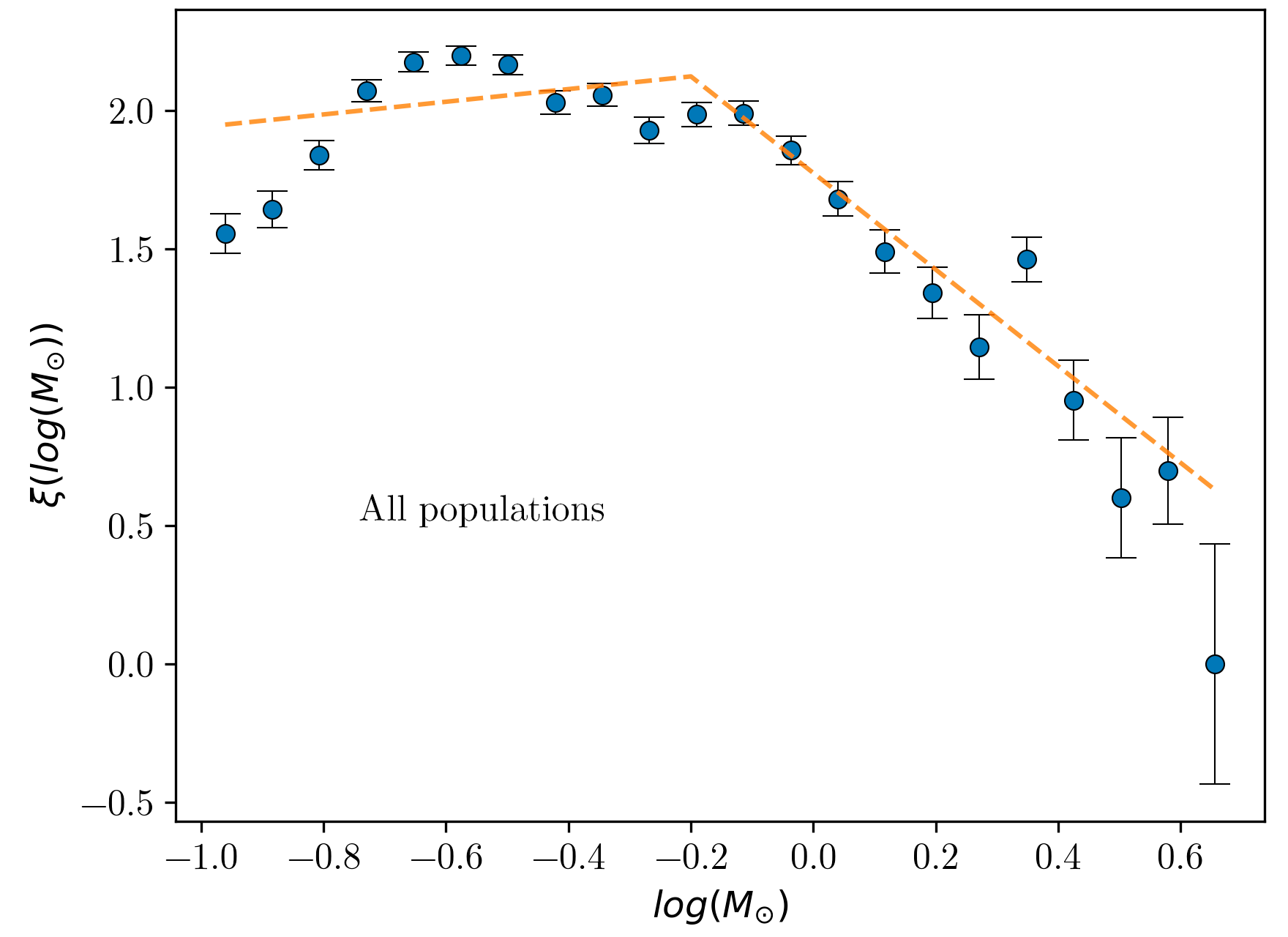}
\caption{Pleiades integrated mass function results for the sample that includes all star populations. The slopes and transition mass point are $\alpha_{A} = (-1.75 \pm 0.33$), $\alpha_{B}= (0.23 \pm 0.22)$ and $M_{c}= (-0.20 \pm 0.07)$.}
\label{pleiades_MF_integrada}
\end{center}
\end{figure}

\begin{figure}
\includegraphics[width=\columnwidth]{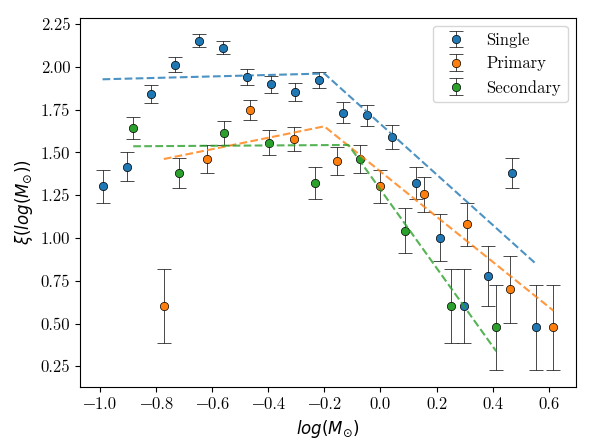}
\caption{Pleiades present-day mass functions for single, primary and secondary stars. The slopes and transition mass points for the single star population are $\alpha_{A}$ = ($-1.99 \pm 0.50$), $\alpha_{B}$ = ($0.03 \pm 0.24$) and $M_{c}$ = ($-0.20 \pm 0.07$), for primary stars $\alpha_{A}$ = ($-1.91 \pm 2.20$), $\alpha_{B}$ = ($0.01 \pm 0.48$) and $M_{c}$ = ($0.20 \pm 0.23$) and the secondary stars $\alpha_{A}$ = ($-2.73 \pm 0.55$), $\alpha_{B}$ = ($0.28 \pm 0.23$) and $M_{c}$ = ($-0.12 \pm 0.05$).}
\label{pleiades_MF_detalhada}
\end{figure}

Applying the two methods described in the previous section, we proceed to obtain the mass of the cluster. From the observed stars we obtain the observed mass and for the unseen stars, we use the mass functions to estimate the unseen mass. In Fig. \ref{pleiades_MF_integrada} we show the present-day mass function obtained from the observed member stars, excluding companion masses as detailed previously for the first method to obtain the mass. From this method, we obtained a mass of ($816 \pm 165$)$M_{\odot}$. For the second method, we used the mass functions shown in Fig. \ref{pleiades_MF_detalhada} where we show the present-day mass functions for single, primary and secondary stars as identified by our method. We performed segmented linear fits for each present-day mass function, where we obtained the slopes and transition mass points for the single, primary and secondary star populations. The results are presented in Table \ref{tab:pleiades-dmf}. From the fits, we obtain the slopes and transition mass points for the single star population which were: $\alpha_{A}$ = ($-1.99 \pm 0.50$), $\alpha_{B}$ = ($0.03 \pm 0.24$) and $M_{c}$ = ($-0.20 \pm 0.07$); for primary stars which were $\alpha_{A}$ = ($-1.91 \pm 2.20$), $\alpha_{B}$ = ($0.01 \pm 0.48$) and $M_{c}$ = ($0.20 \pm 0.23$); and for the secondary stars which were $\alpha_{A}$ = ($-2.73 \pm 0.55$), $\alpha_{B}$ = ($0.28 \pm 0.23$) and $M_{c}$ = ($-0.12 \pm 0.05$). We can see from the results that in general, the slopes for the distinct sections of the mass function agree within the uncertainties obtained. The only exception is the high mass slope of the secondary star population, which is steeper than the others. From these values, we obtain the mass of the cluster using the detailed mass function as described in Sec. \ref{sec:methods}, with a value of ($847 \pm 169$)$M_{\odot}$.

%%%%%%%%%%%%%%%%%%%%%%%%%%%%%%%%%%%%%%%%%%%%%%%%%%%%%%%%%%%%%%%%%%%%%%%%

%%%%%%%%%%%%%%%%%%%%%%%%%%%%%%%%%%%%%%%%%%%%%%%%%%%%%%%%%%%%%%%%%%%%%%%%
%%%%%%%%%%%%%%%%%%%%%%%%%%%%%%%%%%%%%%%%%%%%%%%%%%%%%%%%%%%%%%%%%%%%%%%%

 A summary of our results, with a comparison to other values in the literature, is shown in Table \ref{tabela_Pleiades} where the distinct methods used to estimate the masses are specified. We note that our results show general agreement with other authors, except for the value obtained by  \citet{pleiadesmassatotal1997} which used the tidal radius to estimate the mass.

\begin{table}
\centering
\caption{Parameters of the segmented linear fits obtained for the distinct populations of the Pleiades member stars, present-day mass functions.}
\begin{tabular}{lllll}
\hline
          & \multicolumn{3}{c}{Parameters}                         &  \\
          \hline
          & Low mass         & Transition mass    & High mass        &  \\
          \hline
Single    & $0.03 \pm 0.24$    & $-0.20 \pm 0.07$ & $-1.99 \pm 0.50$  &  \\
Primary   & $0.01 \pm 0.48$  & $0.20 \pm 0.23$    & $-1.91 \pm 2.19$  &  \\
Secondary & $0.28 \pm 0.23$ & $-0.12 \pm 0.05$ & $-2.73 \pm 0.55$ & \\
\hline
\end{tabular}
\label{tab:pleiades-dmf}
\end{table}

\begin{table}
\centering
\caption{Comparison of the mass of Pleiades with the results found in the literature. In the first line, we include the two values determined by us, $(816 \pm 36)M_\odot$ resulting from the integrated mass function and $(838 \pm 21)M_\odot$ from the detailed mass function (MF).}
\begin{tabular}{lcc}
\multicolumn{1}{c}{Reference} & \multicolumn{1}{l}{$M_{T} (m_{\odot})$} & Method \\ 
\hline
This work                                     & $816 \pm 165$  & integrated MF \\
                                              & $847 \pm 169$   & detailed MF \\ 
\citet{pleiadesmassatotal1997}                & 1400          & Tidal radius \\
                                              & 720           & Virial \\
                                              & 950           & integrated MF \\ 
\citet{pleiadesmassatotal1998MNRAS.299..955P} & 735           & King profile \\ 
\citet{pleiadesmassatotal2001AJ....121.2053A} & 800           & King profile \\ 
\citet{massatotalpleiades2008ApJ...678..431C} & $870 \pm 34$  & integrated MF \\ 
\citet{pleiadesmassatotal2019PASP..131d4101G} & $721 \pm 93$  & King profile \\ 
\citet{massatotalrusso2020AstBu..75..407D}    & $855 \pm 104$ & integrated MF \\ 
\citet{masstotalEbrahimi2022}    & $807 \pm 1$ & integrated MF \\ 

\hline
\end{tabular}
\label{tabela_Pleiades}
\end{table}

%%%%%%%%%%%%%%%%%%%%%%%%%%%%%%%%%%%%%%%%%%%%%%%%%%%%%%%%%%%%%%%%%%%%%%%%%%%%%%%%

\section{Mass of real clusters}

In this section, we present the results obtained in our analysis of masses of real open clusters. We have investigated a sample of 1743 open clusters from \citet{catalogoDias2021} for which results of individual stellar membership probabilities as well as fundamental parameters such as age, distance, and extinction were available. For this sample, we updated the memberships and fundamental parameters using data from the \emph{Gaia eDR3} catalogue and obtained the present-day mass functions using the two methods presented in Section \ref{sec:methods}. 

As mentioned before, the final sample was defined with clusters for which the present-day mass function had at least 2 bins on both sides of the transition mass point, a criterion that is sufficient to perform a fit with a straight line. Also, clusters that had visually high star contamination were removed. With this, we were able to obtain the mass for 773 clusters using the integrated mass function and 46 using the detailed mass function. Using this final sample, we investigated relationships with other parameters such as age and size, and location in the Galaxy, among others.

\begin{figure}
\begin{center}
    \includegraphics[width=\columnwidth]{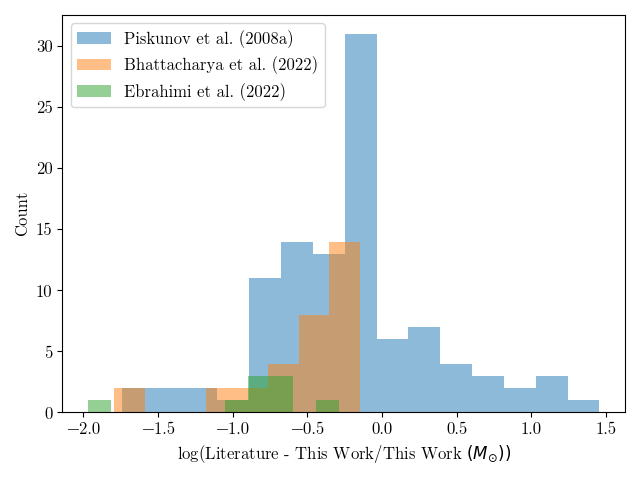}
\end{center}
\caption{Relative error histogram comparing the masses of clusters in common with \citet{PiskunovTidal2008}, \citet{masstotal46ocs2021} and \citet{masstotalEbrahimi2022}.}
\label{fig:allAuthors}
\end{figure}

In Table \ref{tab:totalMasses}, we present an example of the final sample and the masses determined for each cluster, as well as the fundamental parameters adopted for each. The full table is available online.

\begin{table*}
\caption{Sample table of fundamental parameters and masses determined for the open cluster sample using the two methods described in the text. The full table is available online.}
\begin{tabular}{lllllll}
\hline
Name                           & log(age)          & Dist. (kpc)         & Av. (mag.)      & Fe/H                 & $M_{Tint} (m_{\odot})$ & $M_{Tdet} (m_{\odot})$ \\ 
\hline
ASCC 10  &  8.29  $\pm$  0.09  &  0.642  $\pm$  0.003  &  0.8  $\pm$  0.1  &  0.004  $\pm$  0.028  &  229  $\pm$  45  &      \\
ASCC 105  &  8.12  $\pm$  0.05  &  0.552  $\pm$  0.002  &  0.8  $\pm$  0.0  &  0.252  $\pm$  0.066  &  304  $\pm$  60  &   \\
ASCC 107  &  7.65  $\pm$  0.36  &  0.847  $\pm$  0.008  &  1.5  $\pm$  0.2  &  0.253  $\pm$  0.187  &  409  $\pm$  81  &   \\
ASCC 108  &  8.24  $\pm$  0.07  &  1.157  $\pm$  0.003  &  0.7  $\pm$  0.0  &  0.097  $\pm$  0.033  &  935  $\pm$  187  &   \\
ASCC 11  &  8.44  $\pm$  0.19  &  0.826  $\pm$  0.005  &  1.0  $\pm$  0.1  &  0.121  $\pm$  0.044  &  529  $\pm$  105  &  541  $\pm$  108 \\
\hline
\end{tabular}%
\label{tab:totalMasses}
\end{table*}

\subsection{Comparison with literature}

We compared our results of mass of the clusters with those of the large scale published in the literature, briefly summarized as follows and presented in Fig. \ref{fig:allAuthors}.

We found 102 common objects with  \citet{PiskunovTidal2008}, which determined the mass using the King profile. Many objects show significant discrepancies in relation to our results. We find only 30 clusters for which the masses are relatively close to ours, with many objects showing significant discrepancies in relation to our results. The mean of the relative difference between our results was 193\% and the standard deviation was 380\%. The differences can be explained by the different samples of member stars, as well as the different parameters (mainly distance and age) adopted for each cluster.

From the work \cite{masstotal46ocs2021} we found 32 clusters in common which are presented in the histogram of relative errors, as shown in Fig. \ref{fig:allAuthors}. Most clusters are within 60\% of our masses, with a mean relative error of about 35\% and a standard deviation of 20\%. The differences observed are likely due to membership probabilities and fundamental parameters adopted by the authors, which are distinct from the ones used in this work. 

For \cite{Bukowiecki2011b}, which had determination of masses for 599 open clusters, we were not able to find a complete table with the masses for comparison to our results.

The work of \cite{masstotalEbrahimi2022} determined the mass of the Pleiades, which is close to the one found by us (see Table \ref{tabela_Pleiades}). For the other clusters of the work, we found 9 in common and 6 of these are close to our values with a relative error of approximately 15\% - 30\%. Again, differences are likely due to membership probabilities and fundamental parameters adopted by the authors, which are distinct from the ones adopted in this work.

Also, worthy of note, \citet{extrapolacaobr} estimated masses from mass functions obtained using 2MASS photometry. We found 4 clusters in common to their sample and, in general, our results give larger masses with some values agreeing within the uncertainties. For NGC 6694 the authors estimated ($1900 \pm 700$)$M_\odot$ and we obtained ($1208 \pm 242$)$M_\odot$ using integrated mass function method; for NGC 2287 the authors found ($720 \pm 90$)$M_\odot$ and we find ($1185 \pm 237$)$M_\odot$ from the methods detailed mass function and ($1126 \pm 225$)$M_\odot$ integrated MF, respectively; for NGC 2548 and authors estimated a mass of ($630 \pm 150$)$M_\odot$  and our results are ($912 \pm 182$)$M_\odot$ using detailed mass function and ($920 \pm 184$)$M_\odot$ using integrated MF. Finally, for the case of NGC 2682, a mass of ($990 \pm 120$)$M_\odot$ was found by the authors and ($1843\pm 369$)$M_\odot$ according to our results from  integrated mass function method.

Finally, some works have mass estimates for isolated clusters. In the work of \cite{NGC225} two estimates of masses using isochrones fits for NGC 225 were obtained: 1) ($155.2 \pm 2.0$)$M_\odot$ up to magnitude G of 18 of \emph{Gaia EDR3} and assuming a binary fraction of zero; 2) ($170.5 \pm 2.2$)$M_\odot$ with a binary fraction of 0.52. Our results for the integrated mass function returned a mass of ($124 \pm 5$)$M_\odot$ and a fraction of 0.62 binary stars. In \citet{COIN-Gaia} the authors found $439$$M_\odot$ for COIN-Gaia 13 using the integration of the mass function of the observed stars disregarding the influence of binary stars. Regarding this cluster our result was ($367 \pm 9$)$M_\odot$ for detailed mass function and ($378 \pm 16$)$M_\odot$ for integrated MF, we also found a fraction of 0.60 of binaries which can explain part of the difference in masses found.

\subsection{Mass properties of the sample}
\label{sec:totalmasssample}

\begin{figure}
\begin{center}
     \includegraphics[width=\columnwidth]{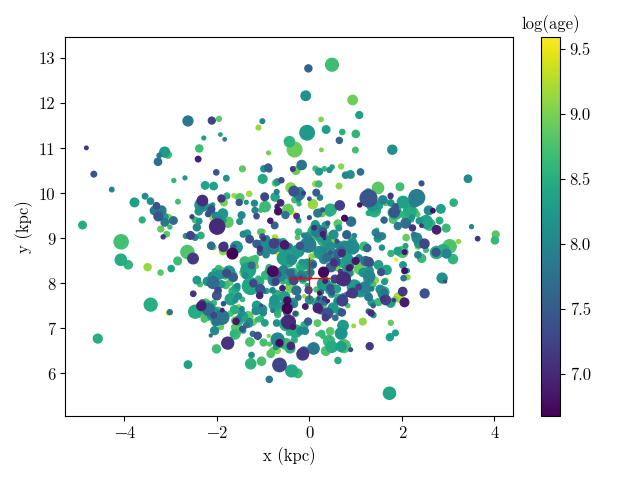}
\caption{Distribution of the 773 open clusters in the galactic plane, with their masses determined using the mass function integrated. The Sun is referenced with a red cross at coordinates (0, 8,3) kpc and the galactic centre is at (0,0). The vector angular velocity is perpendicular to the x-y plane, pointing in the direction of the paper. The size of the dots is proportional to the mass.}
\label{fig:Distocsmt}
\end{center}
\end{figure}

The large number of masses of the open clusters determined in this study allowed us to investigate the
relation of this parameter with others such as age, radius, and mass
segregation, for example. We start by looking at the distribution of open clusters in the Galactic plane, as shown in Fig. \ref{fig:Distocsmt}.  In the figure, the size of the dots is proportional to the mass and no clear pattern of the distribution of massive clusters is seen. 

\begin{figure}
\begin{center}
\includegraphics[width=\columnwidth]{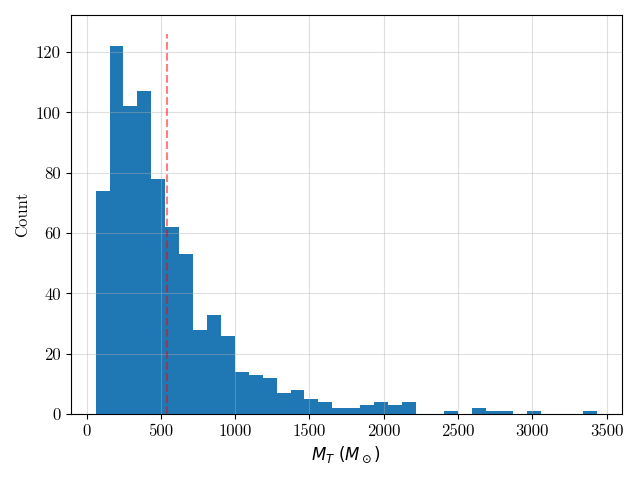}
\caption{Histogram of the mass distribution of clusters in the integrated mass function sample. The vertical line in red indicates the mean of the distribution.}
\label{fig:HistMt}
\end{center}
\end{figure}

In Fig. \ref{fig:HistMt} we show the distribution of cluster masses for our sample. The weighted mean of this distribution is 544$M_{\odot}$ with a standard deviation of 447$M_{\odot}$. 

\begin{figure}
\begin{center}
\includegraphics[width=\columnwidth]{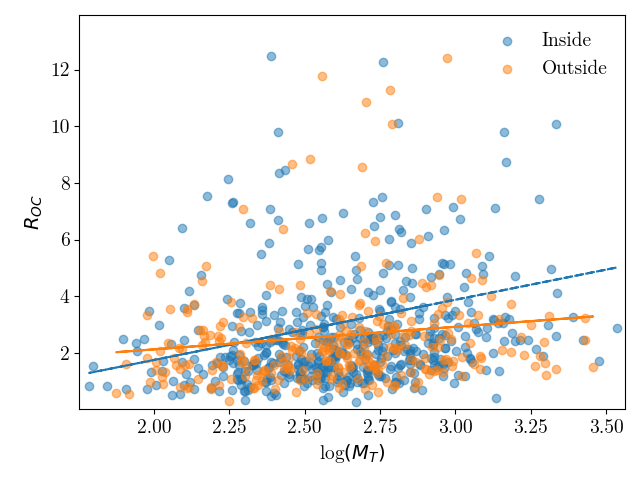}
\caption{Mass-radius distribution for the clusters in the sample of 773 open clusters, with their masses determined by the integrated mass function. The blue dots represent the clusters within the solar orbit, while the orange dots the clusters outside the solar orbit. The lines represent the linear fits for the inner and outer orbit samples, respectively.}
\label{fig:radiusxMt}
\end{center}

\end{figure}

In the work by \citet{Joshi2016}, the authors investigated the relation of the radius and age of the cluster to the mass. The idea is that the mass-radius relationship gives information on the dynamic evolution of the open cluster since, during their early evolution, clusters are expected to experience interactions with massive molecular clouds which can lead to mass loss and a reduction in size. In Fig. \ref{fig:radiusxMt} we show the results we find for the mass and cluster radius relationship. The radii of the clusters adopted are defined as the value that includes 50\% of the stars in the system. For clusters inside the solar orbit we find the correlation $R_{OC} =(2.13\pm1.07)log(M_{T}) - (2.51\pm 2.83)$ and for clusters outside the solar orbit $R_{OC} = (0.80\pm0.96)log(M_{T}) + (0.53\pm 0.96)$. Our results are in agreement with those obtained by \citet{Joshi2016}, where we find few massive clusters with smaller radii as well as a lack of low-mass clusters with large radii. We also see that clusters inside the solar orbit show a steeper correlation, which can be an indication that massive clusters with larger radii have a lower chance of surviving interactions in the direction of the Galactic centre. However, given the uncertainties in the linear fits, the difference is not significant, which is likely due to the large data dispersion.

\begin{figure}
\begin{center}
\includegraphics[width=\columnwidth]{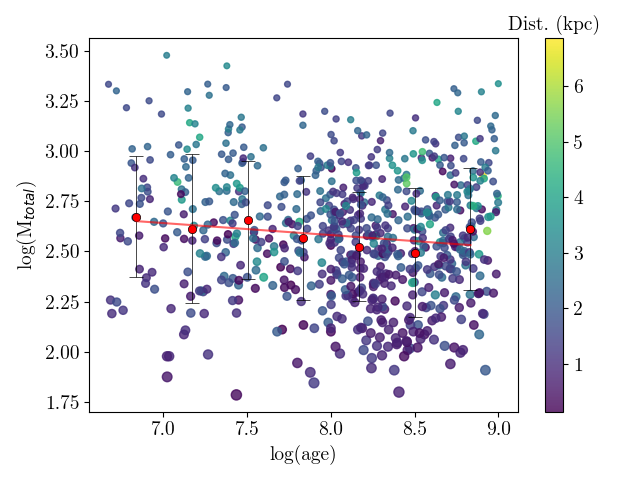}
\caption{Relation of cluster mass to age with red dots representing the average of the masses in a bin in an interval of log(age) = 0.3. Error bars are obtained from the standard deviation of the bin sample.}
\label{fig:AgeXMassMTJosh}
\end{center}
\end{figure}

We have also investigated the mass-age relation for our cluster sample, as shown in Fig. \ref{fig:AgeXMassMTJosh}. Here the general expectation is that interactions with the Galactic tidal field or with the molecular clouds affect their structure and a trend where mass decreases with age should be seen. In our sample, a very tenuous relation is seen, with a correlation coefficient of -0.1, which is considerably lower than the value of -0.95 found by \citet{Joshi2016}. If we take the average of the mass in intervals of log(age) in the same manner as done by \citet{Joshi2016}, excluding clusters with ages greater than log(age) = 9.0 to avoid introducing a selection effect bias, we obtain the following correlation: $log(M_{T}) = (-0.06\pm0.03)log(age) + (3.06\pm 0.24)$. As in \citet{Joshi2016}, we see in our results that younger clusters tend to have higher mass and lose mass as they get older, but the effect is significantly smaller. In, \citet{Joshi2016} the authors found a slope of -0.36 and a mass loss rate $dM_{\odot}/d(age) = 150M_{\odot}$ for the age range 1-10 Myr. In our results we find a mass loss, with a slope of -0.06, that is less intense  and a rate of $dM_{\odot}/d(age) = 13M_{\odot}$ for clusters in the age interval of 1-10 Myr. Here we also performed a LOWESS regression and inspected the uncertainties, finding that a small effect is present, but it is not significant.

\begin{figure}
\begin{center}
\includegraphics[width=\columnwidth]{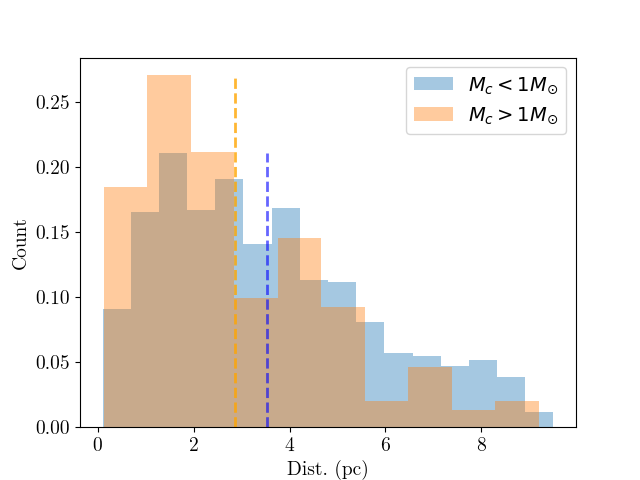}
\caption{Mass segregation for Pleiades, using the distribution of star counts as a function of radius. The member stars were divided into two groups, one with masses $> 1 M_\odot$ and one with $< 1 M_\odot$. The vertical line in orange represents the mean radius of mass concentration for stars $> 1 M_\odot$ and in blue the mean radius for stars $> 1 M_\odot$.}
\label{fig:mass_segregation_pleiades}
\end{center}
\end{figure}

We have also measured the degree of mass segregation within the clusters. To do this, we separate the member stars of a given cluster into two groups: one having members with masses $> 1 M_\odot$ and the other with $< 1 M_\odot$. With these groups, we use two distinct procedures to measure the degree of mass segregation. One, using the ratio of the mean radius of each group's distribution, and the other, using a comparison of the cumulative distribution of mass with respect to the radius, similar to the one used by \citet{MassSeg2022}, where they were concerned in determining whether there was or not mass segregation. Unlike that work, here we quantify the effect using the Kolmogorov-Smirnov measure to quantify the difference between the distributions. As an example, in Fig. \ref{fig:mass_segregation_pleiades} we show the distributions of the two groups as obtained for the Pleiades cluster. In the figure, the difference in the mean radius of each group is indicated and the distinction is clear. Our results corroborate the existence of mass segregation, which was also established by \citet{segregacaopleiades1}, \citet{segregacaopleiades2}, and \citet{Raboud1997} for this cluster.

For the second method, we could determine the significance level of the Kolmogorov-Smirnov measure, indicating how reliable the measured differences between the high and low mass distributions were. With the significance test (p-value) and the adopted threshold of significance of 0.05, we defined a subsample of clusters for which p-value<0.05. We found that in our sample, 368 (48\%) open clusters had a significant level of mass segregation.

\begin{figure}
\begin{center}
\includegraphics[width=\columnwidth]{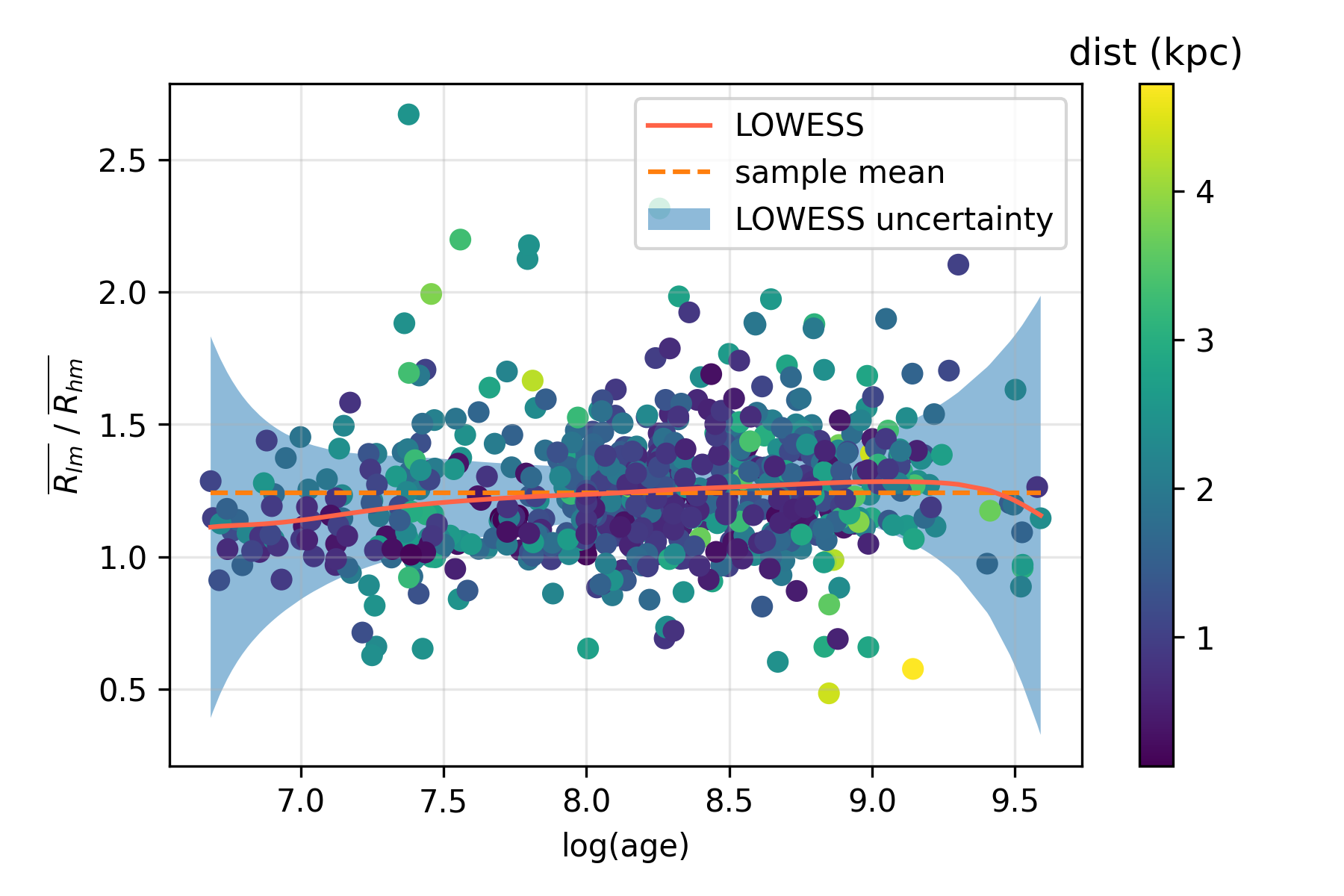} \\
\includegraphics[width=\columnwidth]{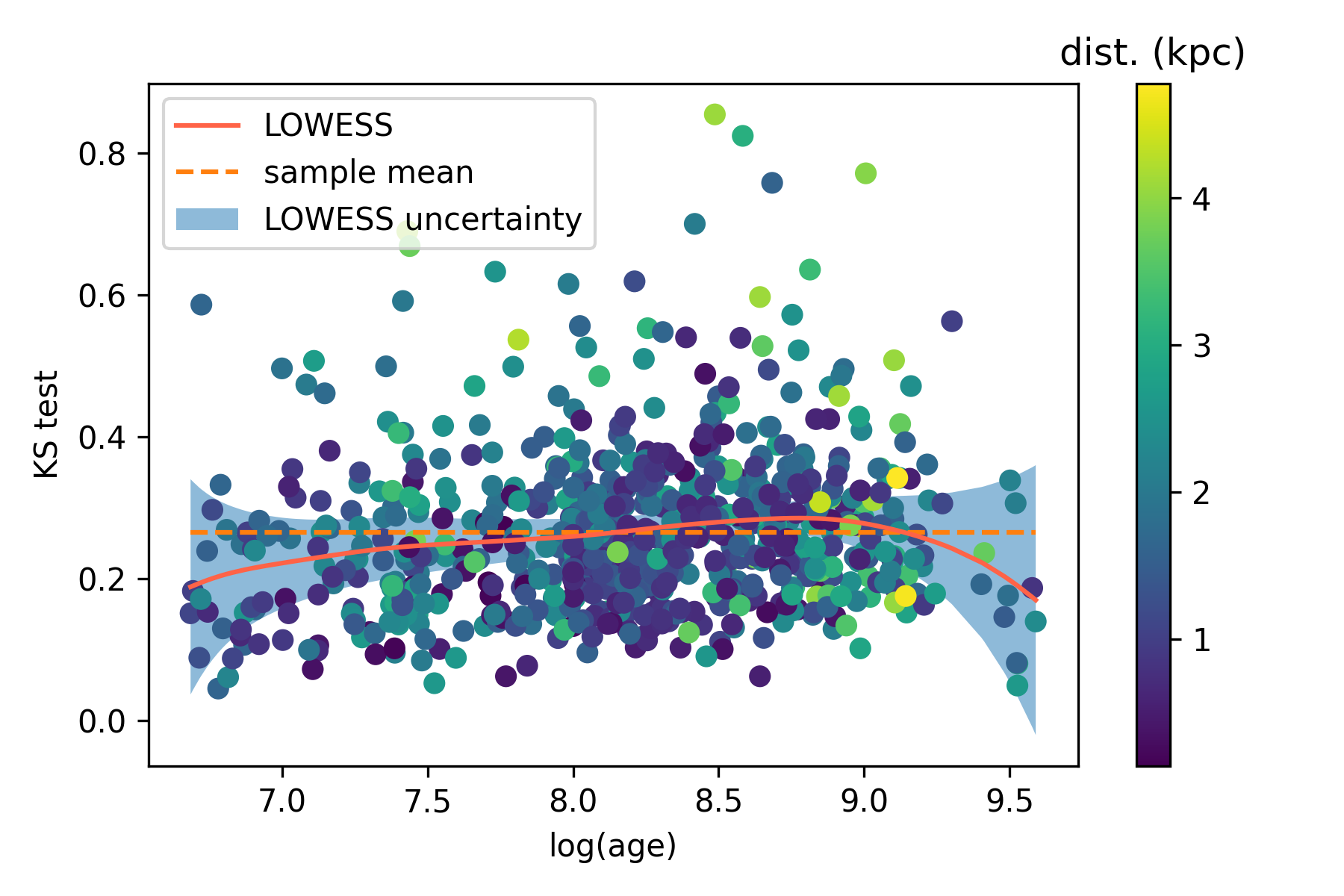}
\caption{Relation of mass segregation to log(age), as measured by the ratio of mean radius for high and low mass groups (upper panel) and Kolmogorov-Smirnov measure (lower panel). The tendency line (red) obtained by LOWESS regression and its uncertainty, as well as the sample mean, is also shown.}
\label{fig:mass_segregation}
\end{center}
\end{figure}

With the mass segregation measure obtained using the two methods, we looked at how it related to the ages of the clusters in our sample. In Fig. \ref{fig:mass_segregation} we show the results for this relation where there is no clear trend in the relation of mass segregation with age. The trend line is obtained with a LOWESS regression procedure where the uncertainties were also estimated, and the sample mean of each segregation statistic is also presented for comparison. In the first panel, we show the result using the first method, based on the ratio of the mean radius of the low mass sample ($\overline{R_{lm}}$) to the mean radius of the high mass sample ($\overline{R_{hm}}$). We see the same behavior for the second method to estimate mass segregation, shown in the second panel of the figure, where the Kolmogorov-Smirnov measure was used. No significant difference between the two methods, as far as the existence of the correlation, is seen. Some authors have mentioned the possibility that low-mass members would tend to escape as the cluster evolves and moves towards a relaxation state. However, we do not detect evidence for that here.

% We determined the following correlation between the ratio and age of clusters: $R = 0.06(\pm 0.01)log(age) + 0.77(\pm 0.12)$. 

\begin{figure}
\begin{center}
\includegraphics[width=\columnwidth]{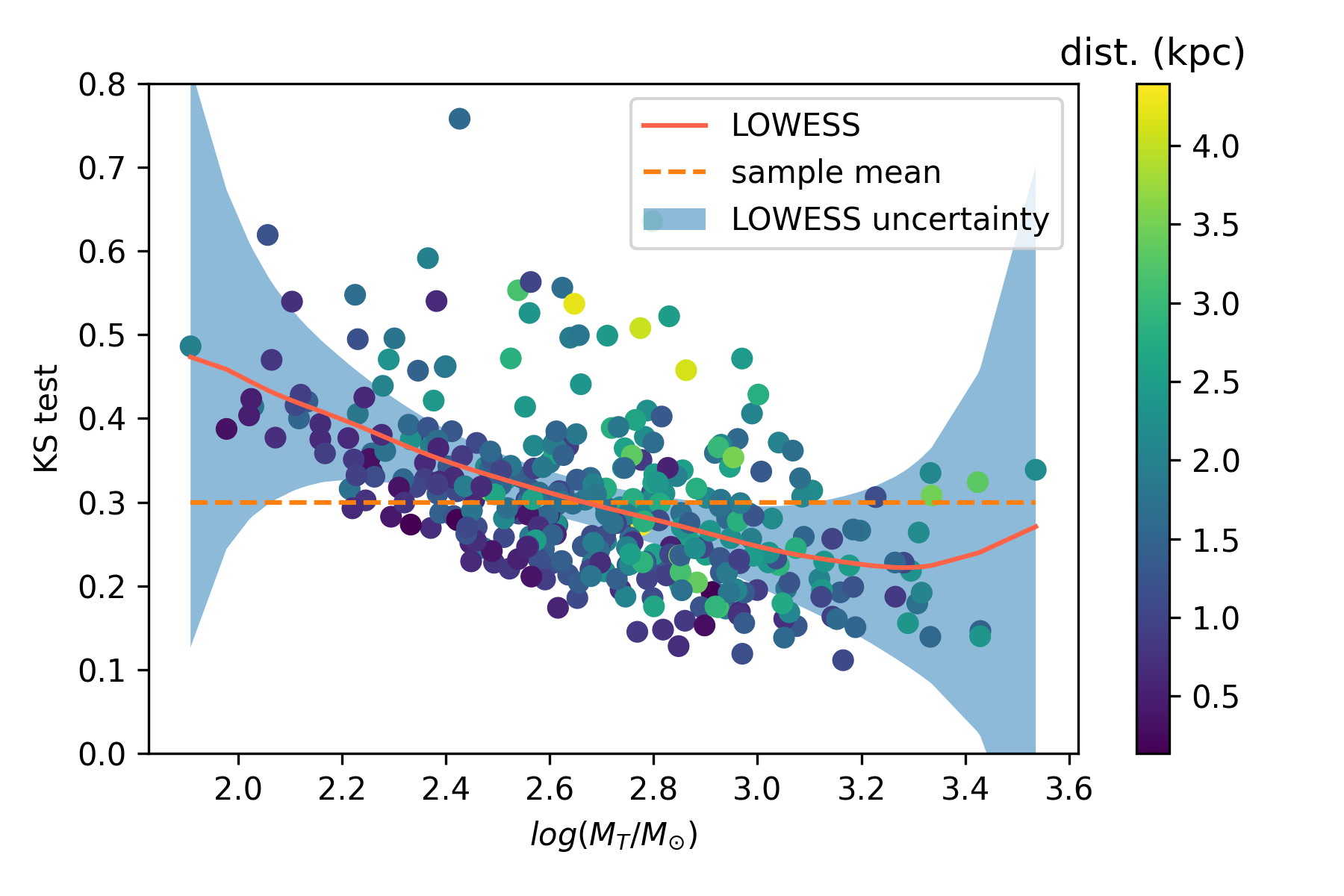}
\caption{Relation of mass segregation, obtained by Kolmogorov-Smirnov measure, to cluster mass for the subsample with the significance level of p<0.05. The tendency line (red) obtained by LOWESS regression and its uncertainty, as well as the sample mean is also shown.}
\label{fig:MtxKStest}
\end{center}
\end{figure}

We also investigated other possible correlations with cluster properties, looking at parameters such as the metallicity, cluster radius, cluster stellar density and cluster mass. We also investigated the sub-sample based on the significance level of the Kolmogorov-Smirnov measure, taking only those clusters for which the distributions showed differences with a p-value lower than 0.05. The only parameter for which a significant correlation was observed was cluster mass. The correlation for this parameter using the high significance sub-sample is shown in Fig. \ref{fig:MtxKStest} where a trend line based on a LOWESS regression is shown with its respective uncertainty.

It is important to note that there is no consensus as to which method to measure mass segregation is ideal. As mentioned in \citet{MassSeg2022}, methods which were considered ideal by \citet{Parker2015}, who compared four different methods which are used to measure mass segregation, were not effective for the clusters they studied. However, both methods adopted in our work give results that agree with each other and, at least statistically in the context of our sample, indicate that mass segregation does not increase with age as has been suggested by other authors.

\section{Conclusions}

In this work, we present a new method for determining individual stellar masses and use it to calculate the masses for a sample of 773 open clusters from the \cite{catalogoDias2021} catalog, with updated memberships and fundamental parameters based on astrometric and photometric data from \emph{Gaia eDR3}.

The method for determining individual masses uses high-quality data of the member stars and fundamental parameters of open clusters, to compare to synthetic clusters generated with full control of relevant parameters, and a Monte-Carlo method to obtain the individual masses and asses its uncertainties and limitations. The masses are estimated by comparing the observed stars with synthetic stars generated in a Monte-Carlo method, finding the nearest star in magnitude space, assigning it a mass, and marking it as binary if applicable. The method was validated with a synthetic cluster grid, and the results were found to be in good agreement with the input masses. Some limitations include difficulty in detecting binary stars in specific regions of the CMD where the main sequence and binary regions overlap, leading to incorrect mass estimation in some cases.

With the individual stellar masses estimated, we determined the mass of the star clusters using two integration methods. In the first method, we integrated the mass function, which contains all the stars in the cluster, a method often used in the literature, and in the second method we performed a detailed integration, which considers a mass function for each population of stars within the cluster (individual, primary and secondary). The unobserved mass was estimated by extrapolating the mass function of the observed stars and accounting for evolved stars and low-mass stars. The method involves obtaining the mass function of the observed star masses and fitting a two-part segmented linear function to it. The contribution of binary systems is estimated by obtaining the mean mass fraction for binary systems and multiplying that value by the estimated mass of single stars when using the integration of the mass function and in detail when using specific population mass functions of the detailed integration method. The method is validated by applying it to a grid of synthetic clusters and comparing the results with the true masses. The results show that we can recover masses with uncertainties under 20\% in most cases, and thus this is the value we adopted as a conservative error estimate.

We applied our methodology to the observed clusters first examining in detail the results for the well-known Pleiades open cluster. The individual masses we estimate for the member stars of this cluster are in agreement with the masses of \citet{GaiaDR3_mass} and the StarHorse catalogue \citep{StarHorse}. The mass obtained for Pleiades using the two methods, the binary fraction and  the average mass ratio of binary systems, showed good agreement with other results found in the literature.

The comparison of our mass estimates of the open clusters to other works in the literature shows good agreement with recent works based on \emph{Gaia} data but large discrepancies with the works of pre-\emph{Gaia} catalogues. Large discrepancies were seen when comparing to the work of \citet{PiskunovTidal2008} who determined masses using the tidal radius, determined from the cluster structural parameters. 

After the thorough validation of the procedures, we focused on the relationship between the mass, cluster radius, and age. The results indicate that the effect seen by \citet{Joshi2016}, that younger clusters tend to have higher mass and lose mass as they get older, is not significant in our sample. We find a mass loss rate of 13$M_{\odot}$/Myr which is significantly lower than the value found by those authors. We find that the correlation between cluster mass and age is weak, with a correlation coefficient of -0.1. The discrepancies in the results are likely due to differences in the samples, memberships and fundamental parameters used.

Looking at the relationship between clusters' masses and radii, we find few massive clusters with smaller radius as well as a lack of low-mass clusters with large radius. The positive correlation between mass and cluster radius could be an indication that clusters inside the solar orbit have a lower chance of surviving interactions in the region of the Galactic centre, however, the difference is not significant. We also looked at the degree of mass segregation within the clusters by measuring it using two distinct methods and found that 368 (48\%) open clusters show a significant level of mass segregation. However, we found no clear significant trend in the relation of mass segregation with age, cluster density, or radius. We did however find a relation between the cluster mass and segregation, showing that more massive clusters tend to be less segregated. Regarding mass segregation, our results are in agreement with \cite{Dib2018}, where no significant relation of mass segregation to age was found.

\section*{Acknowledgements}

Wilton Dias would like to thank CNPq grant 310765/2020-0. Hektor Monteiro would like to thank CNPq grant 436117/2018-5 and FAPEMIG grant APQ 01305-17. This work has made use of the computing facilities available at the Laboratory of Computational Astrophysics of the Universidade Federal de Itajubá (LAC-UNIFEI). The LAC-UNIFEI is maintained with grants from CAPES, CNPq and FAPEMIG.

%%%%%%%%%%%%%%%%%%%%%%%%%%%%%%%%%%%%%%%%%%%%%%%%%%
\section*{Data Availability}

 The data underlying this article, that support the plots and other findings, are available in the article and in its online supplementary material (\url{https://ocmass.streamlit.app}). The updated memberships and individual stellar masses and any other data relevant to the results presented here will be made available upon reasonable requests.

This work has made use of data from the European Space Agency(ESA) {\it Gaia} (http://www.cosmos.esa.int/{\it Gaia}) mission, processed by the {\it Gaia} Data Processing and Analysis Consortium (DPAC,http://www.cosmos.esa.int/web/Gaia/dpac/consortium).

We also employed catalogs from CDS/Simbad (Strasbourg)
and Digitized Sky Survey images from the Space Telescope Science
Institute (US Government grant NAG W-2166)



\bibliographystyle{mnras}
\bibliography{refs} % if your bibtex file is called example.bib

%\begin{thebibliography}{99}
%\end{thebibliography}

%%%%%%%%%%%%%%%%%%%%%%%%%%%%%%%%%%%%%%%%%%%%%%%%%%

%%%%%%%%%%%%%%%%% APPENDICES %%%%%%%%%%%%%%%%%%%%%

\appendix

\section{Online material}

If you want to present additional material which would interrupt the flow of the main paper,
it can be placed in an Appendix which appears after the list of references.

\onecolumn
%\begin{tiny} %\setlength\tabcolsep{12pt}

\begin{longtable}{lllllll}

\caption{Fundamental parameters and masses integrated and detailed determined for the open cluster sample.}
\label{tab:totalMasses-full}\\
\hline
Name & log(age) & Dist. (kpc) & Av. (mag.) & Fe/H                & $M_{Tint} (m_{\odot})$  & $M_{Tdet} (m_{\odot})$ \\ 
\hline 
\endfirsthead
\multicolumn{7}{c}%
{{\bfseries Table \thetable\ continued from previous page}} \\
\hline
Name & log(age) & Dist. (kpc) & Av. (mag.) & FeH                & $M_{Tint} (m_{\odot})$  & $M_{Tdet} (m_{\odot})$ \\
\hline
\endhead
ASCC 10  &  8.292  $\pm$  0.088  &  0.642  $\pm$  0.003  &  0.793  $\pm$  0.068  &  0.004  $\pm$  0.028  &  229  $\pm$  45  &       \\
ASCC 105  &  8.117  $\pm$  0.053  &  0.552  $\pm$  0.002  &  0.765  $\pm$  0.006  &  0.252  $\pm$  0.066  &  304  $\pm$  60  &   \\
ASCC 107  &  7.652  $\pm$  0.359  &  0.847  $\pm$  0.008  &  1.463  $\pm$  0.204  &  0.253  $\pm$  0.187  &  409  $\pm$  81  &   \\
ASCC 108  &  8.241  $\pm$  0.071  &  1.157  $\pm$  0.003  &  0.655  $\pm$  0.050  &  0.097  $\pm$  0.033  &  935  $\pm$  187  &   \\
ASCC 11  &  8.443  $\pm$  0.194  &  0.826  $\pm$  0.005  &  0.975  $\pm$  0.101  &  0.121  $\pm$  0.044  &  529  $\pm$  105  &  541  $\pm$  108 \\
ASCC 113  &  8.493  $\pm$  0.088  &  0.558  $\pm$  0.001  &  0.245  $\pm$  0.049  &  0.001  $\pm$  0.025  &  413  $\pm$  82  &  426  $\pm$  85 \\
ASCC 114  &  8.103  $\pm$  0.089  &  0.900  $\pm$  0.006  &  1.208  $\pm$  0.091  &  0.059  $\pm$  0.055  &  432  $\pm$  86  &   \\
ASCC 115  &  8.308  $\pm$  0.187  &  0.747  $\pm$  0.003  &  0.907  $\pm$  0.090  &  0.049  $\pm$  0.063  &  140  $\pm$  28  &   \\
\hline
\end{longtable}

%%%%%%%%%%%%%%%%%%%%%%%%%%%%%%%%%%%%%%%%%%%%%%%%%%

% Don't change these lines
\bsp	% typesetting comment
\label{lastpage}

\end{document}